\documentclass[prb,twocolumn]{revtex4}%

\usepackage{graphicx}
\usepackage{amsmath}
\usepackage{ulem}
\usepackage{color}

\newcommand{\be}{\begin{equation}}
\newcommand{\ee}{\end{equation}}
\newcommand{\bea}{\begin{eqnarray*}}
\newcommand{\eea}{\end{eqnarray*}}
\newcommand{\bean}{\begin{eqnarray}}
\newcommand{\eean}{\end{eqnarray}}

\begin{document}

\draft
\title
{\bf Large enhancement in thermoelectric efficiency of quantum dot
junction due to increase of level degeneracy}

\author{ David M T Kuo$^{1}$,  Chih-Chieh Chen$^{2}$, and Yia-Chung Chang$^{3,4}$}

\address{$^1$Department of Electrical Engineering and Department of Physics, National Central
University, Chungli, 320 Taiwan}

\address{$^2$Department of Physics, Zhejiang University,  Hangzhou 310027 China}

\address{$^{3}$Research Center for Applied Sciences, Academic Sinica,
Taipei, 11529 Taiwan} \affiliation{$^4$ Department of Physics,
National Cheng Kung University, Tainan, 701 Taiwan}

\date{\today}

\begin{abstract}
It is theoretically demonstrated that the figure of merit ($ZT$)
of quantum dot (QD) junctions can be significantly enhanced when
the degree of degeneracy of the energy levels involved in electron
transport is increased. The theory is based on the the
Green-function approach in the Coulomb blockade regime by
including all correlation functions resulting from
electron-electron interactions associated with the degenerate
levels ($L$). We found that electrical conductance ($G_e$) as well
as electron thermal conductance ($\kappa_e$) are highly dependent
on the level degeneracy ($L$), whereas the Seebeck coefficient
($S$) is not. Therefore, the large enhancement of $ZT$ is mainly
attributed to the increase of $G_e$ when the phonon thermal
conductance ($\kappa_{ph}$) dominates the heat transport of QD
junction system. In the serially coupled double-QD case, we also
obtain a large enhancement of $ZT$ arising from higher $L$. Unlike
$G_e$ and $\kappa_e$, $S$ is found almost independent on electron
inter-dot hopping strength.
\end{abstract}

\maketitle

\section{Introduction}

Recently, many efforts have been devoted to the search of
high-efficiency thermoelectric (TE) materials, because of the high
demand of energy-saving solid state coolers and power
generators.$^{1,2}$ TE devices have very good potential for green
energy applications due to their desirable features, including low
air pollution, low noise, and long operation time. However, there
exists certain barrier for TE devices to replace conventional
refrigerators and power generators since TE materials with figure
of merit ($ZT$) larger than three are not yet found.$^{1,2}$ The
figure of merit, $ZT=S^2 G_e T/\kappa$,defined in the linear
response regime is composed of the Seebeck coefficient ($S$),
electrical conductance ($G_e$), thermal conductance ($\kappa$) and
equilibrium temperature ($T$). $\kappa$ is the sum of the electron
thermal conductance ($\kappa_e$) and phonon thermal conductance
($\kappa_{ph}$). It has been shown that low-dimensional systems
including quantum wells$^{3}$, quantum wires$^{4}$ and quantum
dots (QDs)$^{5}$ have very impressive $ZT$ values when compared
with bulk materials.$^{3-9}$ In particular, $ZT$ of PbSeTe QD
array (QDA) can reach two,$^{5}$ which is mainly attributed to the
reduction of $\kappa_{ph}$ in QDA.$^{2}$ However, QD junctions
with $ZT\ge 3$ are not yet reported experimentally. There are some
technical difficulties in using the QD junction to achieve $ZT\ge
3$ via the reduction of phonon thermal conductivity.$^{1,2}$

More than two decades ago, Hicks and Dresselhauss theoretically
predicted that $ZT$ values of BiTe quantum wells and quantum wires
can be larger than one at room temperature.$^{10,11}$ In
particular, $ZT$ values of nanowires (with diameter smaller than 1
nm) may reach 10 based on the assumption of very low lattice
thermal conductivity ($\kappa_L=1.5W m^{-1}K^{-1}$ for
Bi$_2$Te$_3$). Recently, there are considerable interest on $ZT$
values of nanowires filled with QDs,$^{1,2}$ because it is
expected that $\kappa_{ph}$ can be reduced significantly due to
the introduction of QDs. Such a reduction of $\kappa_{ph}$ due to
phonon scattering with QDs in SiGe nanowire filled with QDs was
verified theoretically in Ref.~12. However, the behaviors of
$G_e$, $S$ and $\kappa_e$ in  nanowires filled with QDs remain
unclear because of the complicated many-body problem involved. The
full many-body effect on the behaviors of electron thermoelectric
coefficients may be analyzed by considering a single QD or double
QDs (DQD) embedded in a single nanowire to reveal the importance
of the electron Coulomb interaction.

Theoretical studies have indicated that a TE device made of
molecular QD junction$^{13-14}$ can reach the Carnot efficiency if
one can neglect $\kappa_{ph}$. Such a divergence of $ZT$ for QDs
is related to the divergence of $G_e/\kappa_e$, which violates the
Wiedeman-Franz law (WFL).$^{15}$ The violation of WFL is a typical
feature for QDs with discrete energy levels.$^{16}$ It is hard to
realize thermal devices with Carnot efficiency as considered in
Refs. 13 and 14, because it is impossible to blockade acoustic
phonon heat flow completely in the implementation of solid state
TE devices$^{1,2}$. Therefore, finding a way to enhance $ZT$ of QD
junctions under an achievable $\kappa_{ph}$ value is crucial.
Here, we demonstrate that by increasing the level degeneracy in
QDs, it is possible to enhance the thermoelectric efficiency
significantly given the condition $\kappa_{ph}/\kappa_e \gg 1$.
The level degeneracy in a QD can be determined by its point-group
symmetry. For spherical QDs made of semiconductors with zincblende
(e.g. III-V compounds) or diamond crystal structure (e.g. Si or
Ge), the point group is Td. Thus, the orbital degeneracy $L$ can
be described by singlet ($A_1$), doublet ($E_2$) or triplet
($T_2$). If the QD energy levels are well described by the
effective-mass model (neglecting the crystal-field effect), then
the orbital degeneracy is determined by the associated orbital
angular momentum quantum number $\ell$, and the level degeneracy
becomes $L=2\ell +1$. For example, the $p$-like states in a
spherical QD are 3-fold degenerate with $L=3$ (not including spin
degeneracy). In an QD junction, one can tune the gate voltage to
access the level with desired degeneracy. The high level
degeneracy ($L$) is also feasible in QDs made of multi-valley
semiconductors such as Si or Ge. Our theoretical results may serve
as useful guideline for optimizing $ZT$ of semiconductor
QD$^{1,2}$ or molecular QD systems$^{13}$, in which a dominating
phonon thermal conductivity cannot be avoided.

\section{Formalism}

Here we consider nanoscale semiconductor QDs embedded in a
nanowire connected with metallic electrodes. An extended Anderson
model is employed to simulate a QD junction with degenerate
levels.$^{17-19}$ The Hamiltonian of the QD junction system
considered is given by $H=H_0+H_{QD}$, where
\begin{eqnarray}
H_0& = &\sum_{k,\sigma} \epsilon_k
a^{\dagger}_{k,\sigma}a_{k,\sigma}+ \sum_{k,\sigma} \epsilon_k
b^{\dagger}_{k,\sigma}b_{k,\sigma}\\ \nonumber
&+&\sum_{k,\ell,\sigma}
V^L_{k,\ell}d^{\dagger}_{\ell,\sigma}a_{k,\sigma}
+\sum_{k,\ell,\sigma}V^R_{k,\ell}d^{\dagger}_{\ell,\sigma}b_{k,\sigma}+c.c.
\end{eqnarray}
The first two terms of Eq.~(1) describe the free electron gas in
the left and right electrodes. $a^{\dagger}_{k,\sigma}$
($b^{\dagger}_{k,\sigma}$) creates  an electron of momentum $k$
and spin $\sigma$ with energy $\epsilon_k$ in the left (right)
electrode. $V^L_{k,\ell}$ ($V^R_{k,\ell}$) describes the coupling
between the the $\ell$-th  energy level of the QD system and left
(right) electrode. $d^{\dagger}_{\ell,\sigma}$ ($d_{\ell,\sigma}$)
creates (destroys) an electron in the $\ell$-th energy level of
the QD.
\begin{small}
\begin{eqnarray}
H_{QD}&=& \sum_{\ell,\sigma} E_{\ell} n_{\ell,\sigma}+
\sum_{\ell} U_{\ell} n_{\ell,\sigma} n_{\ell,\bar\sigma}\\
\nonumber &+&\frac{1}{2}\sum_{\ell,j,\sigma,\sigma'}
U_{\ell,j}n_{\ell,\sigma}n_{j,\sigma'}
\end{eqnarray}
\end{small}
where { $E_{\ell}$} is the spin-independent QD energy level, and
$n_{\ell,\sigma}=d^{\dagger}_{\ell,\sigma}d_{\ell,\sigma}$,
$U_{\ell}$ and $U_{\ell,j}$ describe the intralevel and interlevel
Coulomb interactions, respectively. For nanoscale semiconductor
QDs, the interlevel Coulomb interactions as well as intralevel
Coulomb interactions play a significant role on the electron
transport in semiconductor  junctions. It is worth noting that
$H_{QD}$ possesses the particle-hole symmetry. One can prove it
with a simple swap of electron and hole operators
($d_{\ell,\sigma} \rightarrow c^{\dagger}_{\ell,\sigma}$). The
form of $H_{QD}$ is changed only by constant terms when QD energy
levels are degenerate. This indicates that dynamic physical
quantity is unchanged in the hole picture.

To reveal the transport properties of a QD junction connected with
metallic electrodes, it is convenient to use the Green-function
technique. The electron and heat currents from reservoir $\alpha$
to the QD are calculated according to the Meir-Wingreen
formula$^{19}$
\begin{eqnarray}
J^n_\alpha &=&\frac{ie}{h}\sum_{j\sigma}\int {d\epsilon}(\frac{\epsilon-\mu_{\alpha}}{e})^{n} \Gamma^\alpha_{j} [
G^{<}_{j\sigma} (\epsilon)\\ \nonumber &+& f_\alpha (\epsilon)(
G^{r}_{j\sigma}(\epsilon) - G^{a}_{j\sigma}(\epsilon) ) ],
\end{eqnarray}
where $n=0$ is for the electrical current and $n=1$  for the heat
current.
$\Gamma^\alpha_j(\epsilon)=\sum_{k}|V_{k,j}|^2\delta(\epsilon-\epsilon_k)$
is the tunneling rate for electrons from the $\alpha$-th reservoir
and entering the $j$-th energy level of the QD.
$f_{\alpha}(\epsilon)=1/\{\exp[(\epsilon-\mu_{\alpha})/k_BT_{\alpha}]+1\}$
denotes the Fermi distribution function for the $\alpha$-th
electrode, where $\mu_\alpha$  and $T_{\alpha}$ are the chemical
potential and the temperature of the $\alpha$ electrode. $e$, $h$,
and $k_B$ denote the electron charge, the Planck's constant, and
the Boltzmann constant, respectively. $G^{<}_{j\sigma}
(\epsilon)$, $G^{r}_{j\sigma}(\epsilon)$, and
$G^{a}_{j\sigma}(\epsilon)$ denote the frequency-domain
representations of the one-particle lessor, retarded, and advanced
Green's functions, respectively.
\subsection{Thermoelectric coefficients}
Thermoelectric coefficients including $G_e$, $S$ and $\kappa_e$ in
the linear response regime can be evaluated by using Eq.~(3) with
small $\Delta V=(\mu_L-\mu_R)/e$ and $\Delta T=T_L-T_R$. We obtain
the following expressions of thermoelectric coefficients:
\begin{eqnarray}
G_e&=& (\frac{\delta J^{0}_{\alpha}}{\delta \Delta V})_{\Delta T=0} \\
S&=& - (\frac{\delta J^{0}_{\alpha}}{\delta \Delta T})_{\Delta V=0}/ (\frac{\delta J^{0}_{\alpha}}{\delta \Delta V})_{\Delta T=0} \\
\kappa_e &=& (\frac{\delta J^{1}_{\alpha}}{\delta \Delta
T})_{\Delta
V=0}+(\frac{\delta J^{1}_{\alpha}}{\delta \Delta V })_{\Delta T=0}S\\
&=&(\frac{\delta J^{1}_{\alpha}}{\delta \Delta T})_{\Delta
V=0}-S^2G_eT \nonumber
\end{eqnarray}
where
\begin{eqnarray}
&& (\frac{\delta J^{0}_{\alpha}}{\delta \Delta V})_{\Delta T=0}=\frac{ie}{h}\sum_{j\sigma}\int {d\epsilon}\Gamma^\alpha_{j}(\epsilon)\times \nonumber \\
&& [ \frac{\delta G^{<}_{j\sigma}(\epsilon)}{\delta
f_{\alpha}(\epsilon)} + ( G^{r}_{j\sigma}(\epsilon) -
G^{(a}_{j\sigma}(\epsilon) ) ]\frac{\delta f_{\alpha}(\epsilon)}{\delta \Delta V},
\end{eqnarray}
\begin{eqnarray}
&&(\frac{\delta J^{0}_{\alpha}}{\delta \Delta T})_{\Delta V=0}=\frac{ie}{h}\sum_{j\sigma}\int {d\epsilon}\Gamma^\alpha_{j}(\epsilon)\times \nonumber \\
&& [ \frac{\delta G^{<}_{j\sigma}(\epsilon)}{\delta
f_{\alpha}(\epsilon)} + ( G^{r}_{j\sigma}(\epsilon)  -
G^{(a}_{j\sigma}(\epsilon) ) ]\frac{\delta f_{\alpha}(\epsilon)}{\delta \Delta T},
\end{eqnarray}
\begin{eqnarray}
&&(\frac{\delta J^{1}_{\alpha}}{\delta \Delta T})_{\Delta V=0}=\frac{i}{h}\sum_{j\sigma}\int {d\epsilon}\Gamma^\alpha_{j}(\epsilon)(\epsilon-E_F) \times \nonumber  \\
&& [ \frac{\delta G^{<}_{j\sigma} (\epsilon)}{\delta
f_{\alpha}(\epsilon)} + ( G^{r}_{j\sigma}(\epsilon) -
G^{a}_{j\sigma}(\epsilon) ) ]\frac{\delta f_{\alpha}(\epsilon)
}{\delta \Delta T} ,
\end{eqnarray}
\begin{eqnarray}
&&(\frac{\delta J^{1}_{\alpha}}{\delta \Delta V})_{\Delta T=0}=\frac{i}{h}\sum_{j\sigma}\int {d\epsilon}\Gamma^\alpha_{j}(\epsilon)(\epsilon-E_F) \times  \nonumber \\
&& [ \frac{\delta G^{<}_{j\sigma}(\epsilon)}{\delta
f_{\alpha}(\epsilon)} + ( G^{r}_{j\sigma}(\epsilon) -
G^{a}_{j\sigma}(\epsilon) ) ] \frac{\delta
f_{\alpha}(\epsilon)}{\delta \Delta V}.
\end{eqnarray}
$\frac{\delta G^{<}_{j\sigma}(\epsilon)}{\delta
f_{\alpha}(\epsilon)}$ is obtained by taking the derivative of the
equation of motion with respect to the change in Fermi-Dirac
distribution, $f_{\alpha}(\epsilon)$. Here we have assumed the
variation of the correlation functions with respect to $\delta
f_{\alpha}(\epsilon)$ is of the second order. Note that we have to
take the limit $\Delta V \rightarrow 0$ for the calculation of
$(\frac{\delta J^{0}_{\alpha}}{\delta \Delta V})_{\Delta T=0}$ and
$(\frac{\delta J^{1}_{\alpha}}{\delta \Delta V})_{\Delta T=0}$.
$E_F$ is the Fermi energy of electrodes. The one-particle Green's
functions in Eqs. (7)-(10) are related recursively to high-order
Green's functions and correlation functions via a hierarchy of
equations of motion (EOM)$^{20}$. This hierarchy self terminates
at the $2N$-particle Green function, where $N$ is the number of
levels considered in the QD or coupled QDs.

To reveal the effect of degenerate levels on the thermoelectric
efficiency of QD junction system, all needed Green's functions and
correlation functions arising from electron-electron interactions
in the QDs considered are computed self-consistently following the
procedures described in our previous work.$^{20,21}$ Our procedure
is beyond the mean-field theory, which is widely used in solving
the equation of motion in the Green function calculation.$^{14}$
For $L=4$, our calculation involves solving one-, two-, $\cdots$,
up to eight-particle Green functions.

\subsection{Phonon thermal conductance}
The thermoelectric efficiency of  a QD junction embedded in a
nanowire is determined by the figure of merit,
$ZT=S^2G_eT/(\kappa_e+\kappa_{ph})$, which involves the
$\kappa_{ph}$ of the QD junction system.  The optimization of
molecular QD junctions under the condition of
$\kappa_e/\kappa_{ph}\gg 1$ has been theoretically investigated in
references[13,14]. However, the condition of $\kappa_e/\kappa_{ph}
\gg 1$ is very difficult to realize in practice. The main goal of
this study is to investigate the effect of energy level degeneracy
on thermoelectric efficiency under the realistic condition with
$\kappa_{ph}/\kappa_e > 1$. The phonon thermal conductance of
nanowires have been extensively studied experimentally and
theoretically.$^{22-32}$ In Refs.~22-24 it has been shown
experimentally that $\kappa_{ph}$ displays a linear $T$ behavior
from $20 K$ to $300K$ for silicon nanowires with diameter $22~nm$.
The linear $T$ behavior of $\kappa_{ph}$ also holds for $T$
between $100 K$ and $400 K$ for germanium nanowires with diameter
$19~nm$.$^{25}$ Due to the reduction of $\kappa_{ph}$, $ZT$ of
silicon nanowires increases significantly (with $ZT=1$ at $200 K$)
in comparison with $ZT=0.01$ for bulk silicon at room
temperature.$^{4,26}$  The linear $T$ behavior of nanowires is an
interesting topic. Many theoretical efforts have been devoted to
clarifying why nanowires with diameters near $20~nm$ exhibit the
linear $T$ behavior.$^{27-32}$ For a  true one-dimensional system,
the linear $T$ behavior of $\kappa_{ph}$ is expected.$^{33}$ To
include $\kappa_{ph}$, we have adopted the Landauer formula given
in Refs. 28 and 30.
\begin{equation} \kappa_{ph}=\frac 1 h \int d\omega {\cal T}(\omega)_{ph}
\frac {\hbar^3 \omega^2}{k_B T^2}\frac
{e^{\hbar\omega/k_BT}}{(e^{\hbar\omega/k_BT}-1)^2}, \label{phC}
\end{equation}
where $\omega$ and ${\cal T}_{ph}(\omega)$ are the phonon
frequency and throughput function, respectively. In a perfect wire
throughput is unity for each open channel, then $\kappa_{ph}$ in
such a perfect case is given by
$\kappa_{ph,0}=\frac{k^2_B\pi^2TN_{ph}}{3h}=g_0(T)N_{ph}$, where
$N_{ph}$ is the total number of open modes and
$g_0(T)=\frac{k^2_B\pi^2}{3h}T= 9.456\times 10^{-4}T (nW/K^{2})$
is called the quantum conductance.[34] Experimentally, it was
found that the linear-$T$ behavior in ballistic regime only holds
for temperature below $0.8K$ for wires of size ~200nm with
$N_{ph}= 4$ (which includes one longitudinal, one torsional, and
two flexural modes)$^{34}$. Beyond $0.8K$, the nonlienar-$T$
behavior was observed due the contribution of high-energy phonon
modes being thermally populated. Low-temperature thermoelectric
properties of Kondo insulator nanowire was also studied in Ref. 35
by using the Callaway model to describe $\kappa_{ph}$.

If there exists phonon elastic scattering from the disorder
effects of nanowire surface$^{27-33}$ or interface boundary of QDs
embedded in the nanowire,$^{36,37}$ the throughput function
becomes more complicated.$^{27-33}$ In general, the ${\cal
T}_{ph}(\omega)$ depends on the length ($\tilde L$) and diameter
($D$) of the nanowire, phonon mean free path $\ell_0(\omega)$, and
Debye frequency ($\omega_D$). Realistic calculation of ${\cal
T}_{ph}$ requires heavy numerical work for treating the detailed
phonon dispersion curves,$^{29-33,36-37}$ which is beyond the
scope of this article. However, an empirical expression which
works well in general for semiconductor nanowires at a wide range
of temperature can be found in Ref. 24, which reads
\begin{equation}
{\cal T}_{ph}(\omega)=\frac{N_{ph,1}(\omega)}{1+\tilde
L/\ell_0(\omega)}+\frac{N_{ph,2}(\omega)}{1+\tilde L/D}
\end{equation}
with the frequency-dependent mean free path $\ell_0(\omega)$ given by
\begin{equation}
\frac{1}{\ell_0(\omega)}=B\frac{\delta^2}{D^3}(\frac{\omega}{\omega_D})^2 N_{ph}(\omega),
\end{equation}
where
$N_{ph}(\omega)=4+A(\frac{D}{a})^2(\frac{\omega}{\omega_D})^2$
(for $\omega < \omega_D$) denotes the number of phonon modes. The
dimensionless parameters are chosen to be $A=2.17$ and $B=1.2$.
Notation $a$ denotes the lattice constant of nanowire.
$N_{ph,1}(\omega)=N_{ph}(min(\omega,v_s/\delta))$ and
$N_{ph,2}(\omega)=N_{ph}(\omega)-N_{ph,1}(\omega)$. $v_s$ is the
sound velocity of nanowire and $\delta$ describes the thickness of
the rough surface of nanowire$^{24,28}$ In Eq.~(12), one
essentially replaces the  frequency-dependent mean free path
$\ell_0(\omega)$ by a constant $D$ for the high-frequency modes
($\omega > v_s/\delta$).

For a certain range of temperatures, a simple expression of $\kappa_{ph} (T)$ for molecular QD junction system may be used.
One can approximately write$^{13}$
\begin{equation}
\kappa_{ph}=F_s g_0(T),
\end{equation}
where $F_s$ is a dimensionless correction factor used to describe
the effect of non-ballistic phonon transport due to surface
roughness and phonon scattering from QDs, which replaces the
throughput function ${\cal T}_{ph}(\omega)$ used in Eq.~(11). The
simple expression of Eq.~(14) with $F_s=0.1$ will be used to
describe $\kappa_{ph}$ throughout this article except in Fig.~4.
In Fig.~4, we compare $ZT$ as a function of temperature obtained
by using both Eq.~(12) and Eq.~(14). It is found that with the
simple scaling factor $F_s$ we can describe the behavior of
$\kappa_{ph}$ reasonably well for thin nanowires in the
temperature range of interest. With this simple scaling we can
clarify the effect of level degeneracy ($L$) on $ZT$ for different
magnitudes of $\kappa_{ph}$.

\section{Results and discussion}
Based on Eqs. (4)-(6), we numerically calculate thermoelectric
coefficients including all correlation functions arising from
electron Coulomb interactions in the QDs. Fig. 1 shows $ZT$ of a
single QD junction as a function of the QD level $E_0$, which is
tuned by gate voltage $V_g$ according to $E_0=E_F+50\Gamma_0-eV_g$
for the case of non degeneracy ($L=1$) and 3-fold degeneracy
($L=3$).  Note that the role of gate voltage introduced here allow
us to tune the difference between the QD level energy and Fermi
energy. Throughout this article, we adopt a symmetrical tunneling
rate with $\Gamma_L=\Gamma_R=\Gamma=\Gamma_0$ and all energy
scales are in terms of $\Gamma_0$. $\Gamma_0\approx 1 meV$ in
typically QD junctions; thus, reasonable values for $U_0$ and
$U_I$  in realistic semiconductor QDs are in the range of 20-100
$\Gamma_0$. Fig. 1(a), (b) and (c) are for $k_BT=1\Gamma_0$,
$k_BT=5\Gamma_0$ and $k_BT=10\Gamma_0$, respectively. It is seen
that the maximum $ZT$ for the 3-fold case is significant higher
than the corresponding value for the non-degenerate case when the
temperature is high. For example, the maximum $ZT$ (labeled by
$(ZT)_{max}$) is enhanced by near two-times for $k_BT=5\Gamma_0$
and more than two-times for $k_BT=10\Gamma_0$, although the
enhancement of $ZT$ for $L=3$ is small at $k_B T=1\Gamma_0$. We
observe several new spectral features with similar $ZT$ values at
$E_0$ values spaced apart approximately by the charging energy
$U_0$ or $U_I$, which is caused by the intralevel and interlevel
Coulomb interactions. For the non-degenerate case ($L=1$),
{thermoelectric coefficients can be calculated in terms of the
transmission coefficient ${\cal T}_{LR}(\epsilon)$, which can be
expressed} as
\begin{equation}
\frac {{\cal T}_{LR}(\epsilon)}
{4\Gamma_L\Gamma_R}=\frac{1-N_{-\sigma}}{(\epsilon-E_0)^2+\bar\Gamma^2}
+\frac{N_{-\sigma}}{(\epsilon-E_0-U_0)^2+\bar\Gamma^2},
\end{equation}
where $\bar\Gamma=(\Gamma_L+\Gamma_R)$, and $N_{-\sigma}$ denotes
the single-particle occupation number. Eq. (15) illustrates two
resonant peaks at $\epsilon=E_0$ and $\epsilon=E_0+U_0$ with the
probability weights of ($1-N_{-\sigma}$) and $N_{-\sigma}$,
respectively, which are related to the two $M$-shaped spectral
features in $ZT$ (labeled by $\epsilon_{1,1}$ and
$\epsilon_{1,2}$) with the dip position corresponding to the
resonance energies. (Here the intralevel Coulomb interaction used
is $U_0=60\Gamma_0$) Similarly, for $L=3$ at $k_BT=1\Gamma_0$ we
label the six $M$-shaped spectral features by $\epsilon_{3,n}$
($n=1,\cdots 6)$, which result from the resonant channels at
$E_0$, $E_0+U_{I}$, $E_0+2U_{I}$, $E_0+U_{0}+2U_I$,
$E_0+U_{0}+3U_I$, and $E_0+U_{0}+4U_I$, respectively. (Here, we
have adopted $U_0=U_I=20\Gamma_0$. $U_I$ denotes the interlevel
Coulomb interactions for the $L=3$ case.) These channels
correspond to physical processes of filling the QD with one to six
electrons. At higher temperatures ($k_BT=5\Gamma_0$ and
$k_BT=10\Gamma_0$), the 1st $M$-shaped spectral feature for $ZT$
is broadened and enlarged. (The last M-shaped $ZT$ feature for
$L=3$ is not shown in Fig. 1) For $L=3$, the other spectral
features of $ZT$ (at $\epsilon_{3,n}$; $n=2,3,4,5$) are
suppressed. This is attributed to the significant reduction of
maximum $S^2$ for those channels. Because of electron-hole
symmetry in the system Hamiltonian, {it is expected} that the
spectrum of $ZT$ is symmetrical about the middle point of the
Coulomb gap (MPCG). {For $L=3$, MPCG occurs at
$eV_g=100\Gamma_0$.} Therefore, we only need to focus on the
analysis of $ZT$ optimization near the first spectral feature in
the level-depletion regime, which is defined as the regime when
the average occupation number of the QD summed over spin ($N_t$)
is less than one. In general, it occurs at $E_0 > E_F$.

\begin{figure}[h]
\centering
\includegraphics[angle=-90,scale=0.3]{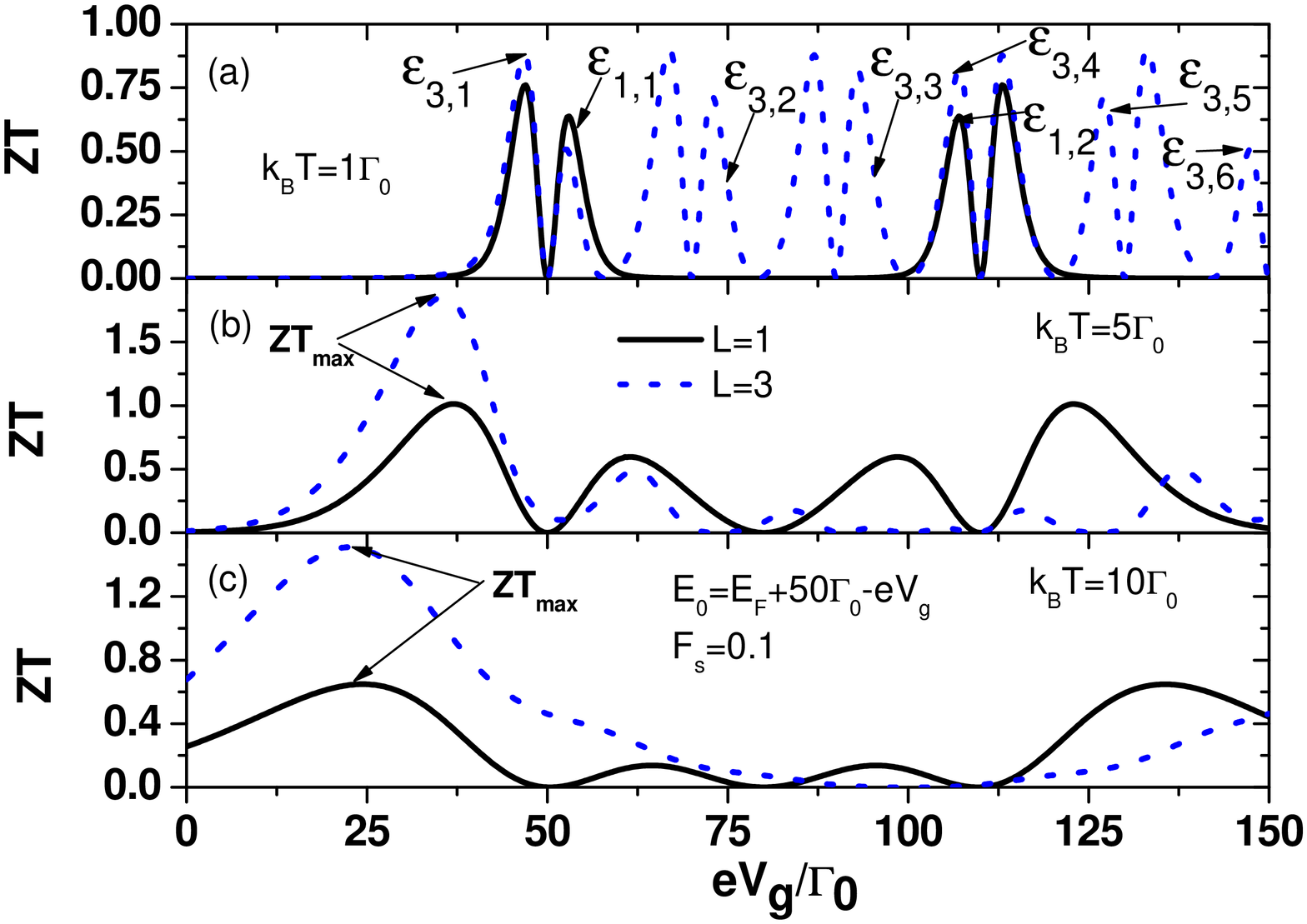}
\caption{Figure of merit as a function of QD energy level tuned by
gate voltage ($E_0=E_F+50\Gamma_0-eV_g$) for level degeneracy,
$L=1$ and 3. (a) $k_BT=1\Gamma_0$, (b) $k_BT=5\Gamma_0$, and (c)
$k_BT=10\Gamma_0$. The correction factor for phonon scattering,
$F_s=0.1$. We have adopted the intralevel Coulomb interaction
$U_{0}=60\Gamma_0$ for $L=1$ and $U_{0}=U_{I}=20\Gamma_0$ for
$L=3$. $U_{I}$ denotes the interlevel Coulomb interaction.}
\end{figure}

To gain better understanding of the enhancement mechanism for
$(ZT)_{max}$ resulting from increased degeneracy, we calculate the
$G_e$, $S$, $\kappa_e$ and $ZT$ of the QD junction as functions of
the level energy ($\Delta=E_0-E_F$) at $k_BT=5\Gamma_0$ for $L=1$
and $L=3$ and the results are shown in Fig.~2. In Fig. 2(a) the
maximum $G_e$ value is enhanced with increasing of degeneracy,
although its dependence of $L$ is not linear.This is mainly
attributed to complicated correlation functions arising from the
electron Coulomb interactions in QD. $G_e$ is much smaller than
$G_0=\frac{2e^2}{h}$ (the electron quantum conductance) even for
L=3, which is mainly attributed to strong electron Coulomb
interactions. We note that the Seebeck coefficient is almost
independent of $L$, whereas $G_e$ and $\kappa_e$ are enhanced with
increasing $L$. However, since $\kappa_{ph}/\kappa_e \gg 1$, the
$L$ dependence of $\kappa_e$ won't affect $ZT$ appreciably. Thus,
the enhancement of $ZT$ shown in Fig. 1(b) mainly comes from the
increase of $G_e$, not $S$. In the Coulomb blockade regime,
$\kappa_e$ and $G_e$ are highly suppressed. Thus, if one can
introduce a mechanism to reduce $\kappa_{ph}$ (with $F_s=0.1$ for
example), then the maximum value of $ZT$ can reach 1 for $L=1$ and
around 2 for $L=3$ as illustrated in Fig. 2~(d).

\begin{figure}[h]
\centering
\includegraphics[angle=-90,scale=0.3]{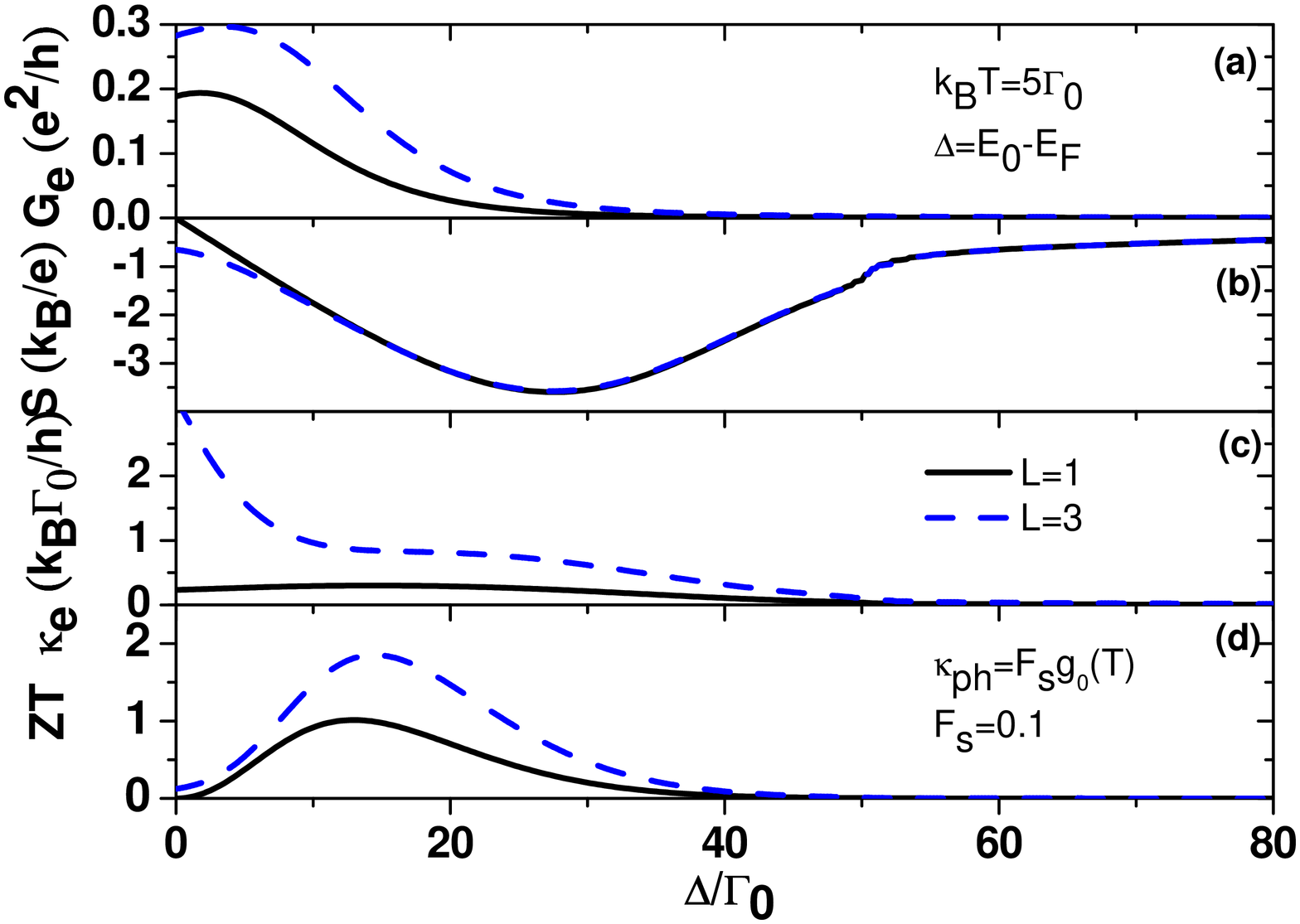}
\caption{(a) Electrical conductance ($G_e$),(b) Seebeck coefficient
(S), (c) electron thermal conductance ($\kappa_e$) and (d) figure of
merit ($ZT$) as a function of QD energy energy level
($\Delta=E_0-E_F$) for different orbital degenerated states at
$k_BT=5\Gamma_0$.  $F_s=0.1$ was used in the calculation of $ZT$.}
\end{figure}

The calculation of thermoelectric coefficients  for the $L=3$ case
including all correlation functions arising from electron Coulomb
interactions is quite complicated. To reveal $L$-dependent
$(ZT)_{max}$, we consider the transmission coefficient ${\cal
T}_{LR}(\epsilon)$ including only the contribution from the
resonant channel $\epsilon-E_0$, which is approximately given by
\begin{equation}{\cal
T}_{LR}(\epsilon)\approx \frac{4\Gamma_L\Gamma_R L
P_{L,1}}{(\epsilon-E_0)^2+\bar\Gamma^2},
\end{equation}
where $P_{L,1}$ is the $L$-dependent probability weight for the
resonant channel at $\epsilon=E_0$. We have
$P_{3,1}=(1-N_{-\sigma})(1-(N_{-\sigma}+N_{\sigma})+c)(1-(N_{-\sigma}+N_{\sigma})+c)$,
where $c=\langle n_{\ell,\sigma}n_{\ell,-\sigma}\rangle$ denotes
the intralevel two particle correlation function.$^{38}$


Thermoelectric coefficients determined by the ${\cal
T}_{LR}(\epsilon)$ of Eqs. (15) and (16) can be calculated by
$G_e=e^2{\cal L}_{0}$, $S=-{\cal L}_{1}/(eT{\cal L}_{0})$ and
$\kappa_e=\frac{1}{T}({\cal L}_{2}-{\cal L}^2_{1}/{\cal L}_{0})$.
${\cal L}_n$ is given by
\begin{equation}
{\cal L}_n=\frac{2}{h}\int d\epsilon {\cal
T}_{LR}(\epsilon)(\epsilon-E_F)^n\frac{\partial
f(\epsilon)}{\partial E_F},
\end{equation}
where $f(\epsilon)=1/(exp^{(\epsilon-E_F)/k_BT}+1)$ is the Fermi
distribution function of electrodes.

Because Eq. (16) does not take into account the interlevel
correlation functions arising from $U_I$, Eq. (16) is not adequate
for illustrating thermoelectric coefficients for the situation
$\Delta/k_BT \le 1$. Nevertheless, we see that $(ZT)_{max}$ of
Fig. 2 does not occur in the $\Delta/k_BT \le 1$ regime.
Therefore, we consider the limit of weak coupling between QD and
electrodes ($\Gamma_L=\Gamma_R=\Gamma \rightarrow 0$) in Eqs. (15)
and (16) and obtain $G_e=\frac{G_0\Gamma \pi
LP_{L,1}}{k_BTcosh^2(\frac{\Delta}{2k_BT})}$, $S=-\Delta/(eT)$,
and $\kappa_e=0$. The $L$-dependent behavior of $(ZT)_{max}$ is
then determined by the simple expression
\begin{equation}
ZT=\frac{(\Delta/eT)^2 G_0 \Gamma \pi L
P_{L,1}T}{k_BTcosh^2(\frac{\Delta}{2k_BT})k_{ph}},
\end{equation}
which explains that the $L$-dependent $ZT$ is determined by $G_e$
rather than $S$ and why $ZT$ approaches 0 as $E_0 \rightarrow E_F$
for $L=1$. Note that $ZT$ for $L=3$ does not approach zero as
$\Delta \rightarrow 0$, because $S$ has a finite value at
$\Delta=0$ (see the dashed line of Fig. 2(b)). Such a result also
indicates that the correlation arising from $U_I$ can not be
neglected for QD when $E_0$ is close to $E_F$. Some novel
nanoscale TE devices resulting from the inclusion of $U_I$ were
theoretically discussed for designing electron heat
rectifiers$^{38}$ and current diodes.$^{39}$

Figure~3 shows $G_e$, $S$, $\kappa_e$ and $ZT$ as functions of
temperature with $\Delta=15\Gamma_0$ for $L=1, 3$, and 4. From the
application point of view, the temperature dependence of $ZT$ is
an important consideration for developing room temperature power
generators used in consumer electronics.$^{2}$ $G_e$ is highly
enhanced for $L=3$ and 4 in the whole temperature regime, but the
difference of $L=3$ and $L=4$ is small, indicating a saturation
behavior as $L$ exceeds 3, mainly because of the factor $P_{L,1}$
in Eq.~(16). As seen in Fig.~3(b), $S$ is nearly independent of
$L$ for $k_BT < 7\Gamma_0$, but becomes weakly dependent on $L$ at
higher temperature. This implies that the effect of resonances at
$\epsilon_{L,2}$ and $\epsilon_{L,3}$ can not be ignored for $L>1$
in the high temperature regime ($ k_BT \ge 7 \Gamma_0$). Although
$\kappa_e$ is enhanced with increasing $L$ as shown in Fig.~3(c),
its effect is insignificant since $\kappa_{e}$ is much smaller
than $\kappa_{ph}$. Therefore, the behavior of $ZT$ with respect
to $k_BT$ is determined by the power factor ($PF=S^2G_eT$). We
found impressive enhancement of $ZT$ for $L=3$ and 4 for $k_BT \ge
4\Gamma_0$. Comparing Figs. 2(d) and 3(d), we see that the maximum
values of $ZT$ occur near $k_BT=\Delta/2.4$ for the tunneling rate
considered ($\Gamma=1\Gamma_0$). Based on such a condition, we can
infer that the maximum $ZT$ at room temperature $k_BT=25~meV$ will
occur near $\Delta=60~meV$ for $\Gamma_0=~1meV$.
\begin{figure}[h]
\centering
\includegraphics[angle=-90,scale=0.3]{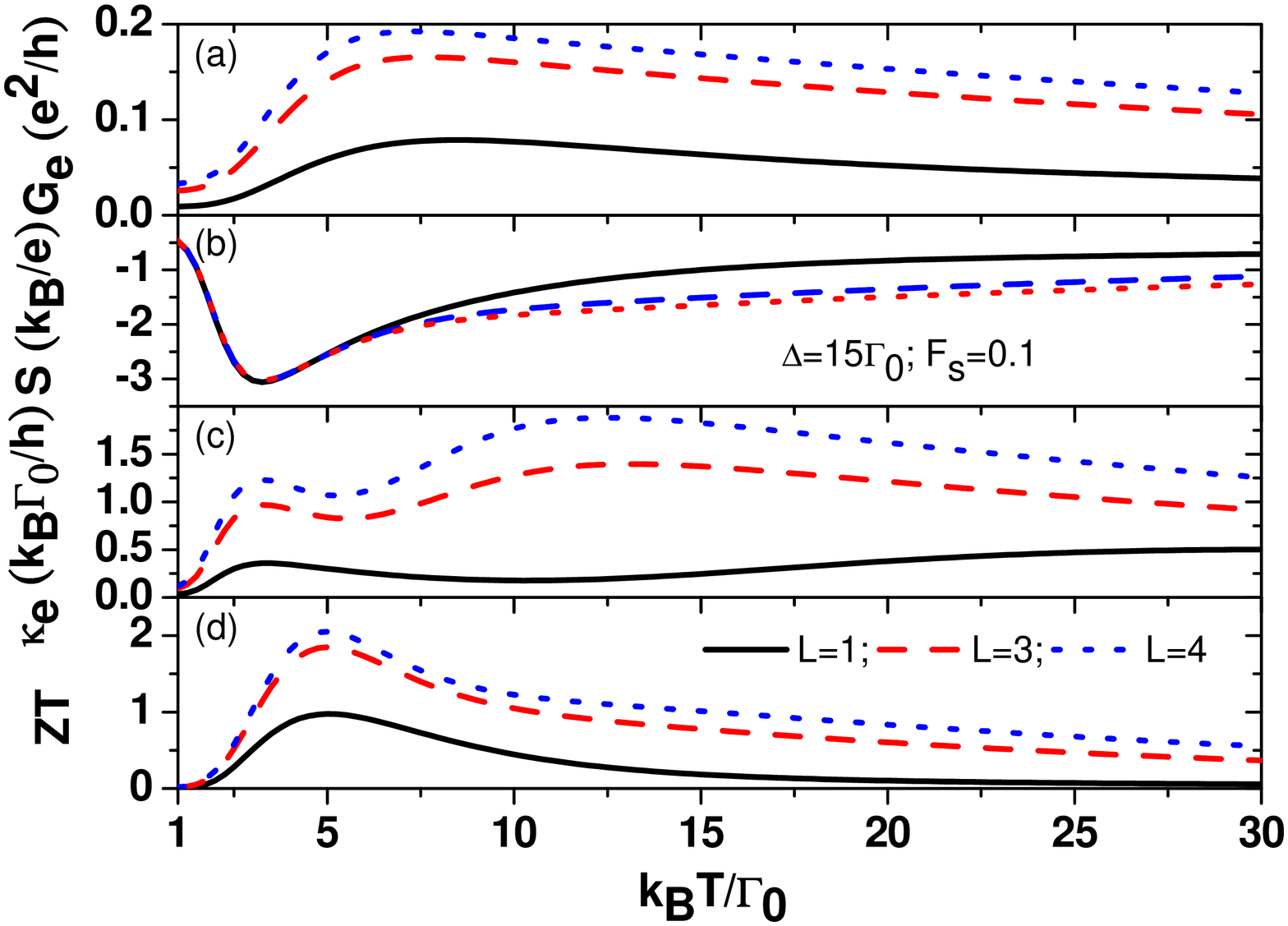}
\caption{(a) Electrical conductance ($G_e$),(b) Seebeck coefficient
(S), (c) electron thermal conductance ($\kappa_e$) and (d) figure of
merit ($ZT$) as a function of $k_BT$ for various values of level degeneracy ($L$)
with $\Delta=15\Gamma_0$. Other physical parameters
are the same as those of Fig. 2.}
\end{figure}

In Figures (1)-(3), $\kappa_{ph}$ is assumed to obey the simple
expression give in Eq. (14). Here we examine whether the large
enhancement of $ZT$ due to level degeneracy will be destroyed when
a more realistic throughput function given by Eq.~(12) is
considered. Fig. 4 shows the comparison of $ZT$ and $\kappa_{ph}$
calculated by both Eq.~(12) and Eq.~(14).  In Fig. 4(a),
$\kappa_{ph}$ used to obtain the solid ($L=3$) and dashed curves
($L=1$) for ZT are calculated by using Eqs.~(11) and (12) with
$D=4~nm$, $\delta=2~nm$ and $\tilde L=2\mu m$. Other parameters
are given by physical properties of silicon
semiconductors.$^{24,28}$ The triangle ($L=3$) and square marks
($L=1$) for ZT are calculated by using $\kappa_{ph}$ based on
Eq.~(14). The maximum $ZT$ values of solid and dashed lines are
near 1.8 and 0.9, respectively. The results of Fig. 4(a) indicate
that the large enhancement of $ZT$ resulting from $L$ is unchanged
even when a more realistic expression for $\kappa_{ph}$ (which is
nonlinear in temperature) is used. When we compare the spectra of
$ZT$ given by the solid curve and the curve with triangle marks
for the case of $L=3$, the curves with triangle marks have better
$ZT$ value at low temperature due to the lower value of
$\kappa_{ph}$ in the linear-$T$ expression. Fig.~4(b) shows
$\kappa_{ph}$ for nanowires with diameter of $10, 15$, and
$20~nm$, which agree well with the experimental results of Ref.
24, lending support for the validity of this model. The comparison
of behaviors of $\kappa_{ph}$ for a 4 nm nanowire obtained by both
Eq.~(12) (solid curve) and Eq.~(14) (triangles) is shown in
Fig.~4(c). It is found that the results obtained by the simple
linear-$T$ expression of Eq.~(14) are fairly close to that
obtained by the  realistic expression of Eq.~(12) for temperatures
between 50K and 200K. In Fig. 4(c), $\kappa_{ph}$ shows a
nonlinear-$T$ behavior between $1K$  and $50$, which is mainly
attributed to frequency-dependent mean free path.  The dashed and
doted lines show the behavior of electronic thermal conductance
($\kappa_e$) with respect to temperature. Note that the $G_e$, $S$
and $\kappa_e$ in Fig.~4 are calculated according to the
simplified method described in Ref.~38, where we only considered
single-particle occupation numbers and intralevel two-particle
correlation functions. The curves with triangle and square marks
shown in Fig.~4(a) are almost identical to the black solid line
and red dashed line of Fig.~3(d) obtained by the full calculation.

\begin{figure}[h]
\centering
\includegraphics[angle=-90,scale=0.3]{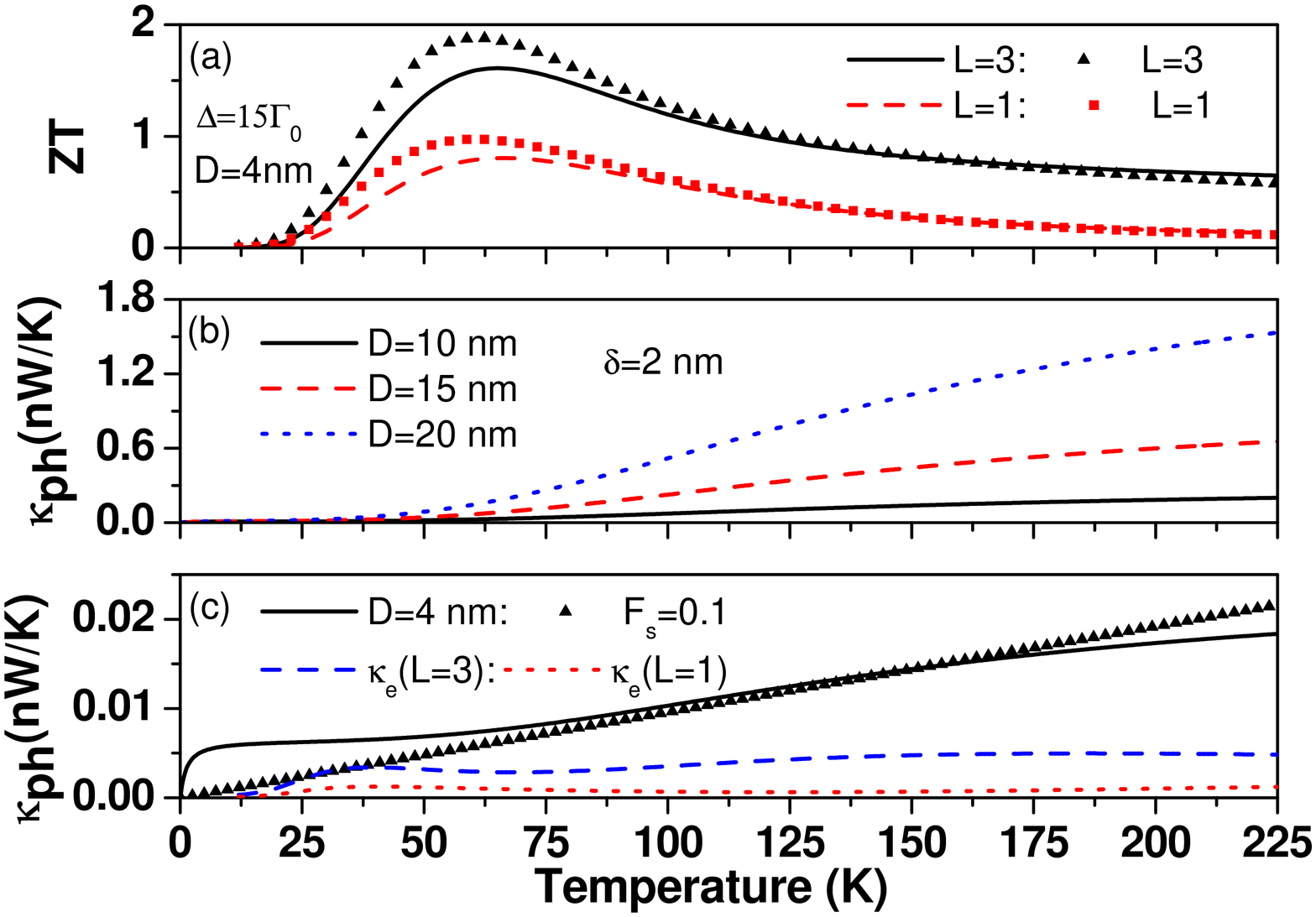}
\caption{(a) Figure of merit ($ZT$) and (b,c) phonon thermal
conductance ($\kappa_{ph}$) as a function of $k_BT$.
The length of nanowire used is $\tilde L=2000~nm$. Other
physical parameters are the same as those of Fig. 3. }
\end{figure}

Next we examine whether the large enhancement of $ZT$ due to
increase of level degeneracy still exists in the case of coupled
double QDs (DQDs). The Hamiltonian of a DQD is given by
$H_{DQD}=H_{QD,L}+H_{QD,R}+U_{LR}\sum_{\ell,j}n_{L,\ell,\sigma}
n_{R,j,\sigma'}+t_{LR}\sum_{\ell,j}
(d^{\dagger}_{L,\ell,\sigma}d_{R,j,\sigma}+h.c)$.$^{40-42}$
$H_{QD,L}$ ($H_{QD,R}$) denotes the Hamiltonian of the left
(right) QD as defined in Eq.~(2). For simplicity, the interdot
electron hopping strengths $(t_{LR})$ and electron Coulomb
interactions ($U_{LR}$) are assumed uniform. Although electron
tunneling currents through DQDs have been extensively studied by
several authors,$^{40-42}$ the optimization of $ZT$ including the
effect of all correlation functions arising from electron Coulomb
interactions has not been reported. Here, we assume one
nondegenerate energy level for each QD ($L=1$). The energy levels
of left QD and right QD are the same (denoted $E_0$). Based on
Eqs.~(4)-(6), $G_e$, $S$, $\kappa_e$ and $ZT$ as functions of the
QD energy level (which is related to the gate voltage by
$E_0=E_F+50\Gamma_0-eV_g$) for $K_BT=3$, 5, and 7 $\Gamma_0$  are
plotted in Fig.~5. There are four peaks labeled by
$\epsilon_{1,n}$ ($n=1,2,3,4$) in the spectrum of electrical
conductance ($G_e$). The Seebeck coefficient ($S$) behaves like
the derivative of $-G_e$ and is vanishingly small at the MPCG due
to electron-hole symmetry. The maximum $S$ occurs near the onset
of the first peak in $G_e$ or the ending of the last peak. Both
$\kappa_e$ and $G_e$ are symmetrical with respect to MPCG. Similar
to the single QD $L=1$ case in Fig.~1(d) the maximum $ZT$ values
occur at either the level-depletion regime or the full-charging
regime as seen in Fig.~5(d).  Here, the maximum $ZT$ is close to
1.7 at $k_BT=3\Gamma_0$ with $F_s=0.1$.

\begin{figure}[h]
\centering
\includegraphics[trim=0cm 0cm 0cm 0cm,clip,angle=-90,scale=0.3]{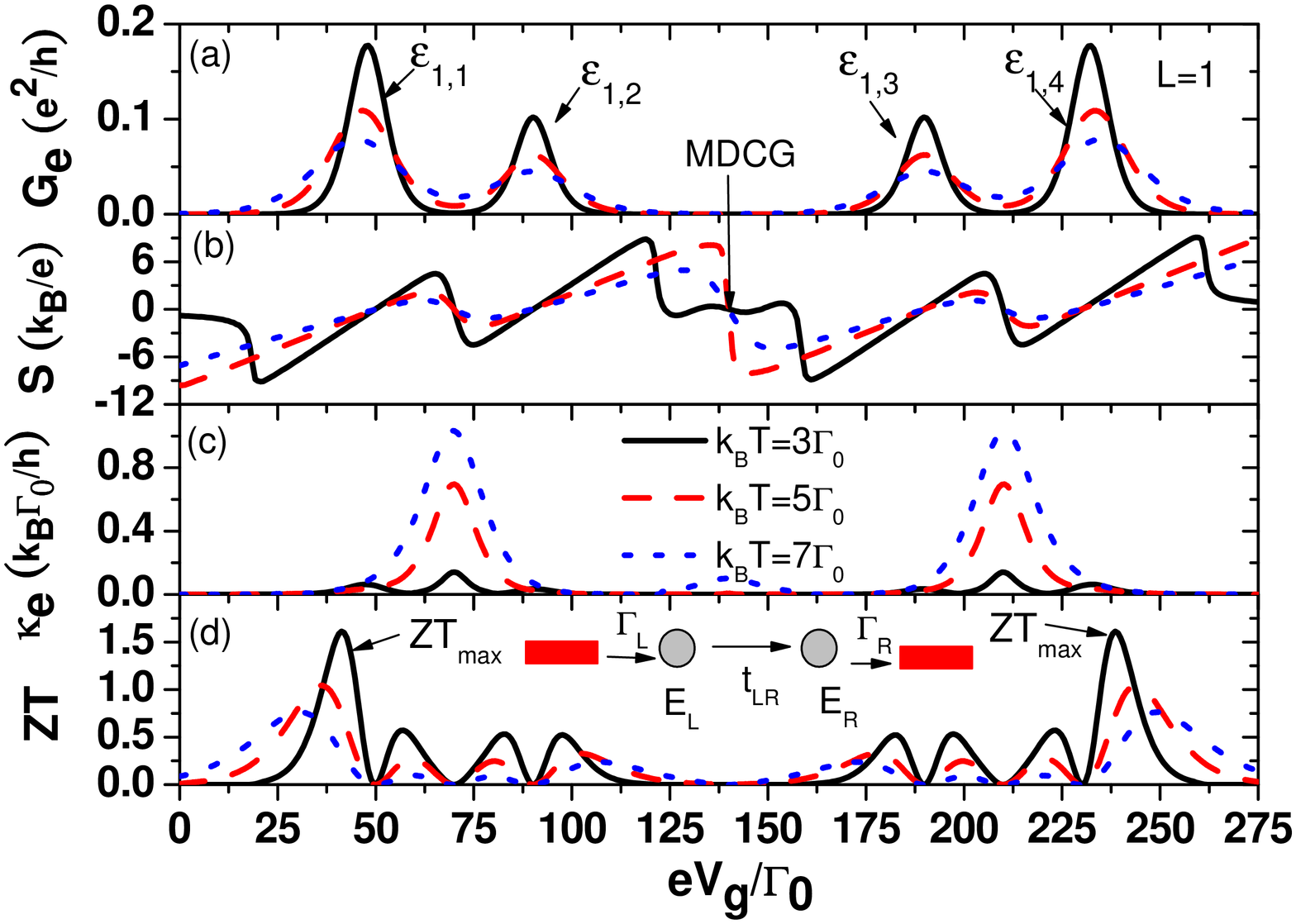}
\caption{(a) Electrical conductance ($G_e$), (b) Seebeck
coefficient ($S$), (c) electron thermal conductance ($\kappa_e$)
and (d) figure of merit ($ZT$) as a function of QD energy level
tuned by gate voltage ($E_0=E_F+50\Gamma_0-eV_g$) in a DQD junction with $L=1$ for various
temperatures. We have considered the electron hopping
strength $t_{LR}=1\Gamma_0$, interdot Coulomb interaction
$U_{LR}=40\Gamma_0$ , intradot Coulomb interactions
$U_{L}=U_{R}=100\Gamma_0$, $\Gamma_L=\Gamma_R=1\Gamma_0$ and
$F_s=0.1$,}
\end{figure}

To further understand the relationship between physical parameters
and thermoelectric coefficients, we consider some approximations
which include only the dominant correlation functions to derive
${\cal T}_{LR}(\epsilon)$ for DQD with L=1. Simple analytic
expressions of
$G^{<}_{j,\sigma}(\epsilon)$,$G^r_{j,\sigma}(\epsilon)$ and
$G^a_{j,\sigma}(\epsilon)$ can be found in our previous
work$^{42}$ when we only include intradot two-particle correlation
functions in the probability weights. Here, we include all
two-particle and three-particle correlation functions. Then, the
following expression of ${\cal T}_{LR}(\epsilon)$ is obtained by
solving the hierarchy of equations of motion (which terminates at
the 4-particle Green function) via a similar procedure as
described in Ref.~42.
\begin{small}
\begin{eqnarray}
& &{\cal T}_{LR}(\epsilon)/(4t^2_{LR}\Gamma_L\Gamma_R)=\frac{P_{1,1} }{|\mu_L\mu_R-t^2_{LR}|^2} \nonumber \\
&+& \frac{P_{1,2} }{|(\mu_L-U_{LR})(\mu_R-U_R)-t^2_{LR}|^2} \nonumber \\
&+& \frac{P_{1,3} }{|(\mu_L-U_{LR})(\mu_R-U_{LR})-t^2_{LR}|^2} \label{TF} \\
\nonumber &+&
\frac{P_{1,4} }{|(\mu_L-2U_{LR})(\mu_R-U_{LR}-U_R)-t^2_{LR}|^2}\\
\nonumber &+& \frac{P_{1,5} }{|(\mu_L-U_{L})(\mu_R-U_{LR})-t^2_{LR}|^2}\\
\nonumber &+& \frac{P_{1,6}
}{|(\mu_L-U_L-U_{LR})(\mu_R-U_R-U_{LR})-t^2_{LR}|^2}\\ \nonumber &+&
\frac{P_{1,7} }{|(\mu_L-U_L-U_{LR})(\mu_R-2U_{LR})-t^2_{LR}|^2}\\
\nonumber
 &+&
\frac{P_{1,8} }{|(\mu_L-U_L-2U_{LR})(\mu_R-U_R-2U_{LR})-t^2_{LR}|^2}, \\
\nonumber
\end{eqnarray}
\end{small}
where $\mu_L=\epsilon-E_L+i\Gamma_L$ and
$\mu_R=\epsilon-E_R+i\Gamma_R$. The probability weights are given
by $P_{1,1}=1-N_{L,\bar\sigma}-N_{R,\bar\sigma}-N_{R,\sigma}+
\langle n_{L,\bar\sigma}n_{R,\bar\sigma}\rangle +\langle
n_{L,\bar\sigma}n_{R,\sigma}\rangle+\langle
n_{R,\bar\sigma}n_{R,\sigma}\rangle-\langle
n_{L,\bar\sigma}n_{R,\bar\sigma}n_{R,\sigma} \rangle$,
$P_{1,2}=N_{R,\bar\sigma}-\langle
n_{L,\bar\sigma}n_{R,\bar\sigma}\rangle -\langle
n_{R,\bar\sigma}n_{R,\sigma}\rangle+\langle
n_{L,\bar\sigma}n_{R,\bar\sigma}n_{R,\sigma} \rangle$,
$P_{1,3}=N_{R,\sigma}-\langle n_{L,\bar\sigma}n_{R,\sigma}\rangle
-\langle n_{R,\bar\sigma}n_{R,\sigma}\rangle+\langle
n_{L,\bar\sigma}n_{R,\bar\sigma}n_{R,\sigma} \rangle$,
$P_{1,4}=\langle n_{R,\bar\sigma}n_{R,\sigma}\rangle-\langle
n_{L,\bar\sigma}n_{R,\bar\sigma}n_{R,\sigma} \rangle$,
$P_{1,5}=N_{L,\bar\sigma}- \langle
n_{L,\bar\sigma}n_{R,\bar\sigma}\rangle -\langle
n_{L,\bar\sigma}n_{R,\sigma}\rangle+\langle
n_{L,\bar\sigma}n_{R,\bar\sigma}n_{R,\sigma} \rangle$,
$P_{1,6}=\langle n_{L,\bar\sigma}n_{R,\bar\sigma}\rangle -\langle
n_{L,\bar\sigma}n_{R,\bar\sigma}n_{R,\sigma} \rangle$,
$P_{1,7}=\langle n_{L,\bar\sigma}n_{R,\sigma}\rangle -\langle
n_{L,\bar\sigma}n_{R,\bar\sigma}n_{R,\sigma} \rangle$, and
$p_{1,8}=\langle n_{L,\bar\sigma}n_{R,\bar\sigma}n_{R,\sigma}
\rangle$, where $\langle n_{\ell,\bar\sigma}n_{j,\sigma}\rangle$
denote the two particle correlation functions and  $\langle
n_{\ell,\bar\sigma}n_{j,\sigma}n_{j,\bar\sigma}\rangle$ the
three-particle correlation functions (including both intradot and
interdot terms). Note that the probability weights satisfy the
conservation law $\sum_{m}P_{1,m}=1$. $U_{L(R)}$ and $U_{L,R}$
denote the intradot and interdot Coulomb interactions. When all
correlation functions of DQDs are included as in Refs. 21 and 40,
it is difficult to find an analytical expression for ${\cal
T}_{LR}(\epsilon)$. The thermoelectric coefficients, $G_e$, $S$,
$\kappa_e$, and $ZT$ obtained by using Eq.~(\ref{TF}) are plotted
in Fig.~6. The results are in very good agreement with those shown
in Fig.~5, which are  obtained by the full calculation, including
all correlation functions. It should be noted that Eq.(\ref{TF})
works well only in the limit $t_{LR}/k_BT\ll 1$. If one would like
to study the spin-dependent thermoelectric coefficients of DQD in
the low temperature regime ($k_BT < t_{LR}$), all correlations
functions should be included.$^{40}$

In Figs. 5 and  6, the peak positions $\epsilon_{1,n}$
($n=1,2,3,4$) correspond to the channels with probability weights
$P_{1,1}$, $P_{1,3}$, $P_{1,6}$ and $P_{1,8}$, respectively. There
exists an interesting behavior for $S$ at low temperature
($k_BT=3\Gamma_0$). We found that $S$ is a linear function of
$eV_g$ near the maximum of $G_e$. From the $\kappa_e$ behavior
shown in  Figs. 5(c) and 6(c), the electron heat flow is maximized
near the mid point between the first (or last) two resonant
channels, and  it increases with increasing temperature.
Eq.~(\ref{TF}) allows us to obtain an analytic form of
thermoelectric coefficients, which is very useful for clarifying
how thermoelectric coefficients are influenced by tunneling rates,
inter-dot hopping strength, and electron Coulomb interactions. Our
analysis shows that $S$ is independent of $t_{LR}$ and it has a
linear dependence of $\Delta=E_0-E_F$ (see Eqs. (21) and (22))
near maximum $ZT$. As a consequence, the trend of $ZT$ with
respect to $t_{LR}$ is determined by that of $G_e$ for
$\kappa_{ph}/\kappa_e \gg 1$.


\begin{figure}[h]
\centering
\includegraphics[angle=-90,scale=0.3]{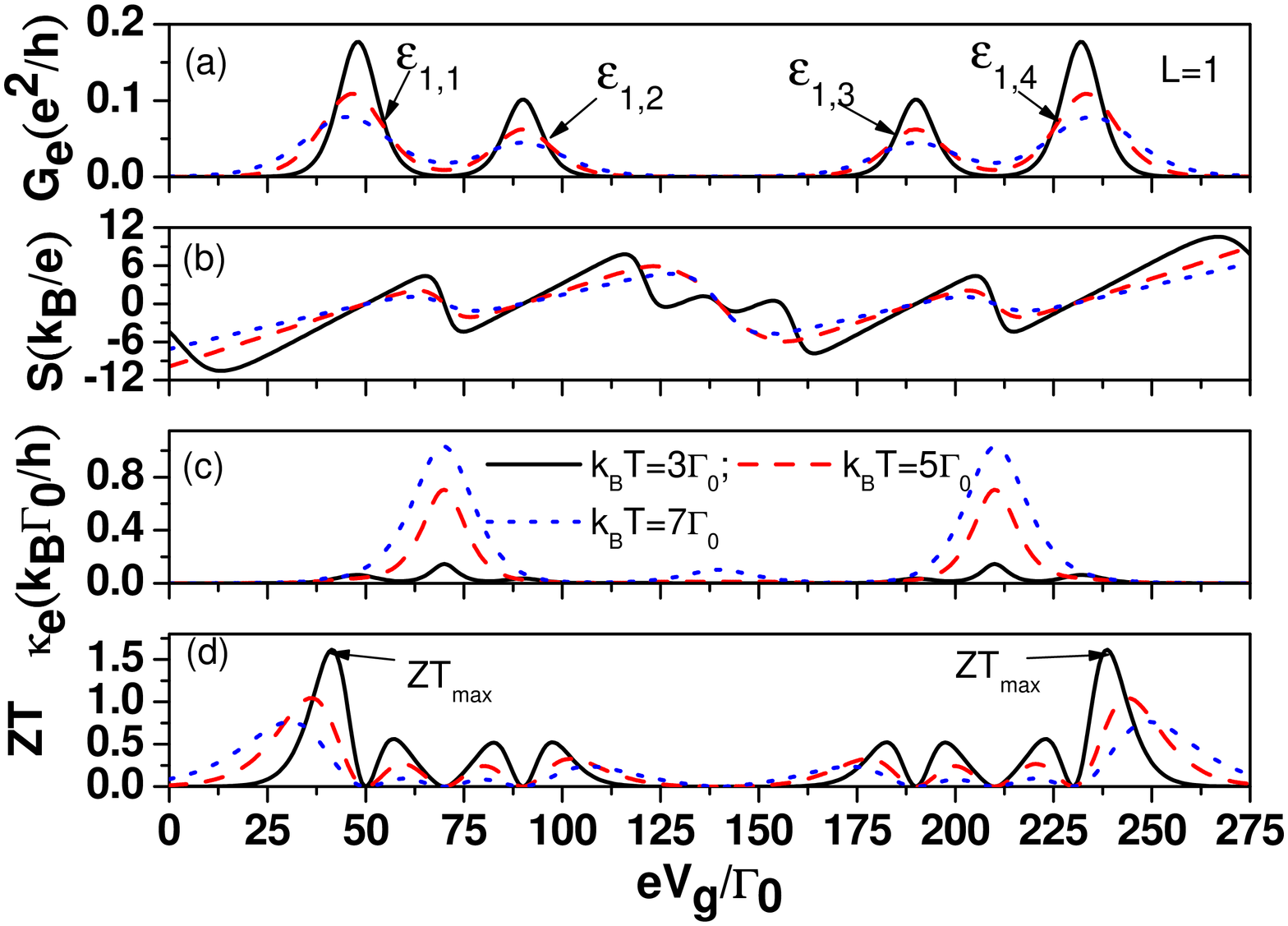}
\caption{  (a) electrical conductance ($G_e$), (b) Seebeck
coefficient ($S$), (c) electron thermal conductance ($\kappa_e$)
and (d) figure of merit ($ZT$) as a function of QD energy level
tuned by gate voltage ($E_0=E_F+50\Gamma_0-eV_g$) in a DQD
junction with $L=1$ for various temperatures calculated by using
Eq.~ (19). Other physical parameters are the same as those of Fig.
5.}
\end{figure}

Next we consider the case with two-fold degenerate levels ($L=2$)
for each QD in the DQD junction. Such two-fold degeneracy can be
realized in QDs with suitable symmetry. For example, the $x$- and
$y$-like states in a disk-shaped QD are degenerate.  Due to
symmetry, the intradot electron hopping process is prohibited,
whereas the interdot electron hopping strength is nonzero. We
assume $t_{LR}=1\Gamma_0$, the same as that of the $L=1$ case. The
intradot Coulomb interactions are taken to be
$U_{L,i,j}=U_{R,i,j}=U_{I}=50\Gamma_0$, where $i,j=1,2$ denote the
two degenerate levels within the same QD. The interdot Coulomb
interaction is taken as $U_{L,R}=40\Gamma_0$. The calculation of
thermoelectric coefficients of DQD with $L=2$ involves solving
one-, two-, $\cdots$, up to eight-particle Green functions. Due to
the presence of $t_{LR}$ term, the numerical procedure is much
more complicated than that of a single QD with $L=4$. Based on
Eqs. (4)-(6), we calculate the thermoelectrical coefficients of
DOD for the $L=2$ as functions of the QD energy level as shown in
Fig.~7. The first four resonant channels of $G_e$ (on the left
hand side of MDCG) are approximately given by
$\epsilon_{2,1}=E_0$, $\epsilon_{2,2}=E_0+U_{LR}$,
$\epsilon_{2,3}=E_0+U_{LR}+U_{I}$, and
$\epsilon_{2,4}=E_0+2U_{LR}+U_{I}$, in which the small $t_{LR}$ is
neglected since $t_{LR}\ll U_{L,R}$. The oscillatory behavior of
$G_e$ displayed in Fig.~7(a) is similar to the $G_e$ spectra
observed experimentally in tunneling current measurements of PbSe
QD (which has a six-fold degenerate excited state) and carbon
nanotube QD (which has an eight-fold state).$^{43,44}$ Although
the $S$ spectrum exhibits more bipolar oscillatory structures, the
maximum $S$ value does not increase with increasing $L$. This
feature is the same as that of a single QD case. Fig.~7(d) shows
an large enhancement of maximum $ZT$ arising from the degeneracy
effect. In the current case, $ZT_{max}$ reaches around 2.7.
Comparing Fig.~7(d) with Fig.~5(d), we see an enhancement of
maximum $ZT$ from around 1.7 to 2.7 when $L$ increases from 1 to 2
for DQD junction when $F_s=0.1$.  We expect even larger
enhancement to occur for higher level degeneracy. Unfortunately,
the computation effort for $L>2$ for a DQD junction is
prohibitively large if all Green functions and correlations
functions are to be included.

To clarify the behavior of $ZT_{max}$ in the level-depletion
regime (with $N_t< 1$), we can approximately write (by keeping
only the dominant channel)
\begin{equation}
{\cal T}_{LR}(\epsilon)\approx\frac{4\Gamma_L\Gamma_R
t^2_{LR}L~P_{L,1}}{|(\epsilon-E_0+i\Gamma_L)
(\epsilon-E_0+i\Gamma_R)-t^2_{LR}|^2},
\end{equation}
where $P_{L,1}$ is the probability weight for DQD in the level-depletion regime.
Under the assumption of $\Gamma_L=\Gamma_R \rightarrow 0$ and
$t_{LR}/k_BT \ll 1$, we have
\begin{equation}
{\cal L}_0=\frac{2}{hk_BT}\frac{\pi\Gamma~t^2_{LR} L
P_{L,1}}{(4t^2_{LR}+\Gamma^2)}
\frac{1}{cosh^2(\frac{\Delta}{2k_BT})}
\end{equation}
and
\begin{equation}
{\cal L}_1=\frac{2}{hk_BT}\frac{\pi\Gamma~t^2_{LR} L
P_{L,1}}{(4t^2_{LR}+\Gamma^2)}
\frac{\Delta}{cosh^2(\frac{\Delta}{2k_BT})}.
\end{equation}
From Eqs. (21) and (22), we have $G_e=e^2{\cal L}_0$ and $
S=-\Delta/(eT)$. This reveals the behavior of $S$ around the
maximum of $G_e$ at $k_BT=3\Gamma_0$ in Fig.~6(b) and L-dependent
$ZT_{max}$ determined by $G_e$ in Fig.~7(d). Note that if we
artificially set $F_s=0$ (i.e. neglecting $\kappa_{ph}$), one can
prove that the enhancement of $ZT_{max}$ arising from $L$ will
disappear due to the $L$-independence of the ratio $G_e/\kappa_e$
and $L$-independence of $S$. If we choose a much higher value of
$F_s$ (e.g. $F_s=1$) in Eq.~(14), we will have the condition
$\kappa_{ph}/\kappa_e \gg 1 $. In this situation, $L$ dependence
of $ZT_{max}$ is fully determined by $G_e$ and we have $ZT$
linearly proportional to $L$, since $P_{L,1}$ in Eq.~(18) is close
to 1 under the condition $(E_0-E_F)/k_BT \gg 1$, and the $ZT$
values will be approximately proportional to $1/F_s$. Finally, we
would like to point out that QD junctions embedded in a silicon
nanowire can be realized by the advanced technique reported in
Refs. 45 and 46.

\begin{figure}[h]
\centering
\includegraphics[angle=-90,scale=0.3]{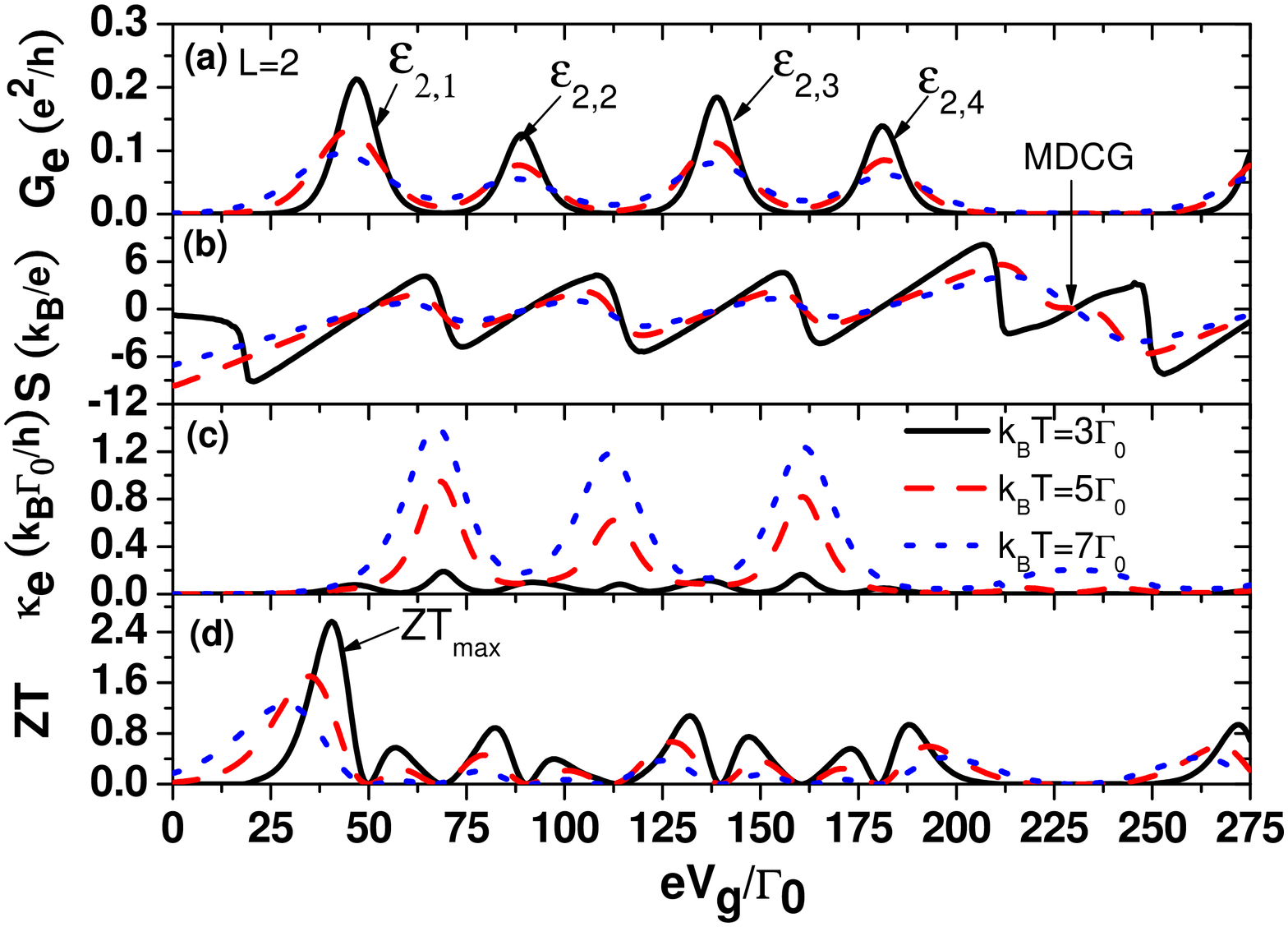}
\caption{(a) Electrical conductance ($G_e$), (b) Seebeck
coefficient ($S$), (c) electron thermal conductance ($\kappa_e$)
and (d) figure of merit ($ZT$) as a function of QD energy level
tuned by gate voltage ($E_0=E_F+50\Gamma_0-eV_g$) in a DQD
junction with $L=2$ for various temperatures.
$U_{L,\ell,j}=U_{R,\ell,j}=50\Gamma_0$ and $U_{LR}=40\Gamma_0$.
Other physical parameters are the same as those of Fig. 5.}
\end{figure}

\section{Conclusion}
We have theoretically investigated the effects of level degeneracy
on thermoelectric properties of QDs embedded in a thin nanowire
junction in the Coulomb blockade regime. All the correlation
functions arising from electron Coulomb interactions for electrons
in the degenerate levels are included in our calculation. We found
that the maximum values of $ZT$ can be highly enhanced with level
degeneracy under the typical condition with $\kappa_{ph}$ much
larger than $\kappa_e$.  When $(E_0-E_F)/k_BT \gg 1 $, $S$ is
independent on $L$. Therefore, the enhancement of $ZT_{max}$ in
the level-depletion regime is mainly attributed to the increase of
$G_e$. Large enhancement of $ZT$ due to the increase of level
degeneracy is also found in the presence of finite electron
hopping in coupled QD system. In our studies, we assumed a simple
expression $\kappa_{ph}=F_s g_0(T)$ for the phonon thermal
conductance. However, it is worth pointing out that our conclusion
on the effect of level degeneracy on $ZT$ is not limited to the
linear $T$-behavior of $\kappa_{ph}$ (as illustrated in Fig. 4).
The enhancement due to level degeneracy holds as long as
$\kappa_{ph} > \kappa_{e}$, regardless of the temperature
dependance of $\kappa_{ph}$. This implies that the design
principle based increasing level degeneracy is applicable for a
composite materials involving arrays of coupled QDs$^{1,2}$ and
molecular QDs.$^{13}$


{\bf Acknowledgments}\\
This work was supported under Contract Nos. MOST 103-2112-M-008-009-MY3 and MOST 104-2112-M-001-009-MY2.
\mbox{}\\
E-mail address: mtkuo@ee.ncu.edu.tw\\
E-mail address: yiachang@gate.sinica.edu.tw\\

\setcounter{section}{0}

\renewcommand{\theequation}{\mbox{A.\arabic{equation}}} 
\setcounter{equation}{0} 

\mbox{}\\


\end{document}